\documentclass[aps,prd,twocolumn,showpacs,showkeys,nofootinbib]{revtex4}
\pdfoutput=1
\usepackage{epsfig}
\usepackage{amsmath}
\usepackage{graphicx}
 \usepackage{multirow}
\newcommand{\beq}{\begin{equation}}
\newcommand{\eeq}{\end{equation}}
\newcommand{\bqa}{\begin{eqnarray}}
\newcommand{\eqa}{\end{eqnarray}}

\newcommand{\bold}{\textbf}

\def\bfsigma{\mbox{\boldmath $\sigma$}}

\begin{document}

\title{Relativistic corrections to the form factors of $B_c$ into $S$-wave charmonium}
\author{Ruilin Zhu}\email{rlzhu@njnu.edu.cn}
\author{Yan Ma }
\author{Xin-Ling Han}
\author{Zhen-Jun Xiao}\email{xiaozhenjun@njnu.edu.cn}
\affiliation{
 Department of Physics and Institute of Theoretical Physics,
Nanjing Normal University, Nanjing, Jiangsu 210023, China\\
 }

\begin{abstract}
We investigate the form factors of $B_c$ meson into $S$-wave charmonium within the
nonrelativistic QCD  effective theory and obtain the next-to-leading order relativistic corrections
to the form factors, where both the $B_c$ meson and the charmonium are treated as
the nonrelativistic bound states. Treating the charm quark as  a light quark in the limit \
$m_c/m_b\to 0$, some form factors are identical at the maximum recoil point, which are consistent
with the predictions in the heavy-quark effective theory and the large-energy effective theory.
Considering that the branching ratios of $B^+_c \to J/\psi D_s^+$ and $B^+_c \to J/\psi D_s^{*+}$ have
been measured by the LHCb and ATLAS Collaborations recently, we employ the form factors of
$B_c$ meson into $S$-wave charmonium at the next-to-leading order  accuracy to these two decay
channels and obtain  more precise predictions of  their decay rates.
Numerical results indicate that the factorizable diagrams dominate the contribution
in these two channels, while the colour-suppressed and the annihilation diagrams
contribute less than 10 percent. Our results are consistent with the LHCb and ATLAS data.

\pacs{12.38.Bx, 14.40.Pq,  12.39.Jh}

\keywords{Perturbative calculations,
Heavy quarkonia,
Nonrelativistic quark model}

\end{abstract}

\maketitle

\section{Introduction}

The Large Hadron Collider (LHC) provides a  solid platform to test the consistency and the correctness
of the Quantum  Chromodynamics (QCD) as the fundamental theory of the strong interaction. On heavy flavor
side, nonrelativistic  QCD (NRQCD) effective theory is a powerful framework to calculate the production
cross section and the decay  rate of heavy quarkonium~\cite{Bodwin:1994jh}.
Because the heavy quark relative velocity $v$ is small in the rest frame of heavy quarkonium,
the cross-sections and the decay rates can be expanded as the series of the  NRQCD long-distance
matrix elements (LDMEs) with the corresponding short-distance coefficients.

The $B^-_c$ meson is composed of two different heavy flavors and has three kinds of decay modes:
(i) the bottom quark decays through $b\to c, u$;
(ii) the charm quark decays through $\bar{c}\to \bar{s}, \bar{d}$;
and (iii) the weak annihilation.
The contributions  to the total decay width of the $B^-_c$ meson are found to be around 20, 70,
and 10 percent for these three categories of decay modes, respectively~\cite{Kar:2013fna}.
Therein the transition of the bottom quark into the charm quark, where the antiquark $\bar{c}$ is the spectator,
has attracted a lot of attentions in both theoretical and experimental communities~\cite{Olive:2016xmw}.

In theoretical side,  the form factors of $B_c$ meson into $S$-wave charmonium have been investigated in
different methods. These methods can be assigned according to the following four types.
(i) The perturbative QCD (PQCD) approach in which the form factors can be computed in terms of the
perturbative hard kernels  and the nonperturbative meson wave functions with the harmonic oscillator
form~\cite{Du:1988ws} or the transverse momentum dependent form~\cite{Sun:2008ew,Wen-Fei:2013uea,Rui:2014tpa}.
(ii) QCD and light-cone sum rules. In QCD sum rules (QCD SR) the form factors are related to the three-point
Green functions~\cite{Colangelo:1992cx,Kiselev:1993ea,Kiselev:1999sc}, while in light-cone sum rules (LCSR)
the form factors depend mainly on the leading twist light cone distribution amplitudes of the mesons~\cite{Huang:2007kb}.
(iii) The relativistic,  nonrelativistic and the light-front quark models.
In relativistic  quark model (RQM), the bottom and charm quarks in mesons are treated as relativistic
objects~\cite{Nobes:2000pm,Ebert:2003cn,Ivanov:2005fd}.
In  the nonrelativistic limit, the form factors have been calculated in a nonrelativistic constituent
quark model (NCQM)~\cite{Hernandez:2006gt}. While in the light-front quark model (LFQM), the form factors
can be extracted from the plus component of the current operator matrix elements~\cite{Wang:2008xt,Ke:2013yka}.
(iV) The NRQCD approach. In this effective theory, the leading order results of the
form factors have been given in Refs.~\cite{Chang:1992pt,Bell:2005gw}.
The next-to-leading order (NLO) corrections have been presented in
Refs.~\cite{Bell:2005gw,Qiao:2011yz,Qiao:2012vt}, in which the dependence of the form factors on the
momentum transfer squared $q^2$ are also obtained. Besides, the optimal renormalization scale of
form factors has also been discussed by using the principle of maximum conformality (PMC) in
Ref.~\cite{Shen:2014msa}. But the form factors have the singularity at the minimum recoil point,
which makes the prediction of form factors in the minimum recoil region invalid~\cite{Qiao:2012vt,Qiao:2012hp}.

On the other hand, some hot topics are studied in heavy flavor field along with continuous accumulation
of experimental data at LHC, which are very helpful to  investigate the properties of form factors.
According to the naive factorization scheme, the decay rates of $B_c$ exclusive decays to a charmonium
and  a light meson such as $B_c \to J/\psi+\pi$ and $B_c \to J/\psi+\rho$ depend mainly on the form
factors at the maximum recoil point. While the decay rates of  $B_c$ semileptonic decays to a charmonium
such as $B_c \to J/\psi+\ell+\bar{\nu}_\ell$  will depend on the form factors in different momentum recoil
region, and the exclusive two-body decays to a charmonium and  a heavy meson such as $B_c \to J/\psi+D_s^{(*)}$
will depend on the form factors far from the maximum recoil point. Therein the  decay channels
$B^+_c \to J/\psi+D_s^{+}$  and $B^+_c \to J/\psi+D_s^{*+}$ were first observed by the LHCb
Collaboration in 2013~\cite{Aaij:2013gia}. These two channels have been studied recently by the
ATLAS Collaboration using a dataset corresponding to integrated luminosities of 4.9$fb^{-1}$ and
$20.6 fb^{-1}$ of $pp$ collisions collected at centre-of-mass energies $\sqrt{s}=$ 7TeV and 8 TeV,
respectively~\cite{Aad:2015eza}. The data indicate some discrepancies from some theoretical
predictions~\cite{Aad:2015eza}. The studies of these channels shall test the form factors'
predictions in different approaches, and  improve our understanding of the nonperturbative properties of QCD.

The paper is organized as the following. In Sec.~\ref{II}, we will give the results of the relativistic
corrections to the form factors of $B_c$ into $S$-wave charmonium within NRQCD approach.
The form factors will be investigated  in the limit of  heavy bottom quark, i.e. $m_c/m_b\to 0$.
In Sec.~\ref{III}, we are going to calculate the decay rates of $B_c \to J/\psi+D_s^{(*)}$ .
In this section, the contributions to the  branching ratios of $B_c \to J/\psi+D_s^{(*)}$ from
the factorizable diagrams, the colour-suppressed and the annihilation diagrams will be considered, respectively.
We summarize and conclude in the end.

\section{Relativistic corrections to the form factors\label{II}}
\subsection{NRQCD approach}

The heavy quark pair inside  the heavy quarkonium is nonrelativistic in the rest frame of heavy
quarkonium, since the heavy quark's mass is much larger than the QCD binding energy. The quark
relative velocity  squared is estimated as $v^2\approx 0.3$ for $J/\psi$ and $v^2\approx 0.1$
for $\Upsilon$~\cite{Bodwin:1994jh}.
If a heavy quarkonium is produced in a hard-scattering process or the heavy quark decays in a
heavy quarkonium, the cross sections and the decay rates can be ordered in powers of both the
strong coupling constant $\alpha_s$ and the  quark relative velocity $v$, which have been investigated in NRQCD effective theory
by Bodwin, Braaten, and Lepage~\cite{Bodwin:1994jh}.

The NRQCD Lagrangian can be written into the following terms~\cite{Bodwin:1994jh}
\begin{eqnarray}
{\mathcal L}_{\rm NRQCD} &=&
\psi^\dagger \left( i D_t + {{\bf D}^2 \over 2m} \right) \psi
+ {c_F \over 2 m} \psi^\dagger \bfsigma \cdot g_s {\bf B} \psi
\nonumber\\
&+& \psi^\dagger {{\bf D}^4 \over 8m^3} \psi+{c_D\over 8 m^2} \psi^\dagger ({\bf D}\cdot g_s {\bf E}- g_s {\bf E}\cdot {\bf D})\psi
\nonumber\\
&+&{i c_S\over 8 m^2} \psi^\dagger \bfsigma \cdot ({\bf D}\times g_s {\bf E}- g_s {\bf E}\times {\bf D})\psi
\nonumber\\
&+& \left(\psi \rightarrow i \sigma ^2 \chi^*, A_\mu \rightarrow - A_\mu^T\right) +
{\mathcal L}_{\rm light} \,,
\label{NRQCD:Lag}
\end{eqnarray}
where ${\mathcal L}_{\rm light}$ represents the Lagrangian for the light quarks and gluons.
$\psi$ and $\chi$ denote the Pauli spinor field that annihilates a heavy quark and creates a heavy antiquark, respectively.
The short-distance coefficients $c_D$, $c_F$, and $c_S$ can be perturbatively expanded in powers
of $\alpha_s$, which can be expressed as $c_i=1+{\cal O}(\alpha_s)$.

The inclusive annihilation decay width of a heavy quarkonium can be factorized as~\cite{Bodwin:1994jh}
\begin{eqnarray}
\Gamma(H)=\sum_n\frac{2\mathrm{Im}f_n(\mu_\Lambda)}{m_Q^{d_n-4}}\langle H|{\cal O}_n(\mu_\Lambda)|H\rangle\,,
\end{eqnarray}
where$\langle H|{\cal O}_n(\mu_\Lambda)|H\rangle$  are  NRQCD annihilation LDMEs, which involve nonperturbative information
 and are ordered by the  relative velocity $v$ between the heavy quark and antiquark
inside the heavy quarkonium $H$.  The heavy quark's mass is denoted as $m_Q$.
The imaginary part of the short-distance coefficients $f_n(\mu_\Lambda)$  can be calculated order by order in the perturbative theory.

The leading order  NRQCD operators for the decay of $S$-wave heavy quarkonium  are
\begin{eqnarray}
\mathcal{O}(^{1}S_{0}^{[1]})&=&\psi^{\dagger}\chi\chi^{\dagger}\psi,\\
\mathcal{O}(^{3}S_{0}^{[1]})&=&\psi^{\dagger}\bfsigma\chi\cdot\chi^{\dagger}\bfsigma\psi.
\end{eqnarray}

The NLO  relativistic correction operators for $S$-wave heavy quarkonium  are
\begin{eqnarray}
\mathcal{P}(^{1}S_{0}^{[1]})&=&\frac{1}{2}\left[\psi^{\dagger}\chi\cdot\chi^{\dagger}(-\frac{i}{2}{\overleftrightarrow{ {\bold D}}})^2\psi+h.c.\right],\\
\mathcal{P}(^{3}S_{1}^{[1]})&=&\frac{1}{2}\left[\psi^{\dagger}\bfsigma\chi\cdot\chi^{\dagger}\bfsigma(-\frac{i}{2}{\overleftrightarrow{ {\bold D}}})^2\psi+h.c.\right],
\end{eqnarray}
where the $h.c.$ denotes the corresponding complex conjugate term. Using the vacuum-saturation approximation,
the  NRQCD LDMEs can be estimated as $\langle H|\mathcal{O}_n| H\rangle\simeq \langle
H|\psi^{\dagger}\mathcal{K}^\prime_n\chi|0\rangle\langle 0|\chi^{\dagger}\mathcal{K}_n\psi| H\rangle$
with $\mathcal{O}_n=\psi^{\dagger}\mathcal{K}^\prime_n\chi\chi^{\dagger}\mathcal{K}_n\psi$.
Furthermore, the vacuum expectations of production operators $\mathcal{O}^H_n$ are related to the decay
matrix elements as $\langle 0|\mathcal{O}^H_n| 0\rangle\simeq (2J+1)\langle H|\mathcal{O}_n| H\rangle$
with heavy quarkonium angular momentum $J$.

\subsection{Covariant projection method}

Instead of  the traditional matching method where both of the QCD and NRQCD calculations are required in order
to extract the short distance coefficients, we will use an equivalent method, i.e.  the covariant projection
method. In order to get the coefficients of the relativistic correction operators, the quark relative momentum
should be kept.  Let $p_1$ and $p_2$ represent the momenta of the heavy quark $Q$ and anti-quark $\bar{Q^\prime}$,
respectively. Without loss of generality, one may decompose the momenta as the following
\begin{eqnarray}
p_1 &=&  \alpha \,P_{H}-k,\\\
p_2 &=&  \beta\, P_{H}+k,
\end{eqnarray}
where $P_{H}$ is the  momentum of the heavy quarkonium, and $k$ is a half of the relative momentum between the
quark pair with $P_{H}\cdot k=0$.  The energy fractions for  $Q$ and $\bar{Q^\prime}$ in heavy quarkonium are
denoted as $\alpha$ and $\beta$, respectively. The explicit expressions for all the momenta in the rest frame
of the  heavy quarkonium are given by
\begin{eqnarray}
P_{H}^\mu &=&  (E_1+E_2,0),\\
k^\mu &=&  (0,\bold{k}\,),\\
p_1^\mu &=&  (E_1,-\bold{k}\,),\\
p_2^\mu &=&  (E_2,\bold{k}\,).
\end{eqnarray}
The heavy quarkonium momentum becomes purely timelike while the relative momentum is   spacelike in the rest frame. $\alpha=E_1/(E_1+E_2)$  and $\beta=1-\alpha$ with the on-shell conditions $E_1=\sqrt{m_1^2-k^2}$, $E_2=\sqrt{m_2^2-k^2}$, and $k^2=-\bold{k}^2$. $m_1$ and $m_2$ denote the masses  of the heavy quark $Q$ and anti-quark $\bar{Q^\prime}$, respectively.

The Dirac spinors for the heavy quark $Q$ and anti-quark $\bar{Q^\prime}$ can be written as
\begin{eqnarray}
u_1(p_1,\lambda) &=&  \sqrt{\frac{E_1+m_1}{2E_1}}\left(
                                           \begin{array}{ll}
                             ~~~~\xi_\lambda \\
\frac{\vec{\sigma}\cdot \overrightarrow{p_1}}{E_1+m_1}\xi_\lambda
                                           \end{array}
                                         \right)\,,
\end{eqnarray}
\begin{eqnarray}
v_2(p_2,\lambda) &=&  \sqrt{\frac{E_2+m_2}{2E_2}}\left(
                                           \begin{array}{ll}
\frac{\vec{\sigma}\cdot \overrightarrow{p_2}}{E_2+m_2}\xi_\lambda\\
                             ~~~~\xi_\lambda\end{array}
                                         \right)\,,
\end{eqnarray}
where $\xi_\lambda$ is the two-component Pauli spinors and $\lambda$ is the polarization quantum number.

One can easily get the covariant expressions for the spin-singlet and spin-triplet combinations
of spinor bilinearities.  The corresponding projection operators are
\begin{widetext}
\begin{eqnarray}
\Pi_S(k) &=&  -i\sum_{\lambda_1,\lambda_2} u_b(p_1,\lambda_1)\bar{v}_c(p_2,\lambda_2)\langle\frac{1}{2}\lambda_1\frac{1}{2}\lambda_2|S S_z\rangle\otimes \frac{\bold{1}_c}{\sqrt{N_c}}\nonumber\\
&=&\frac{i}{4\sqrt{2 E_1 E_2}\omega}(\alpha \,p\!\!\!\slash_{H}-k\!\!\!\slash+m_1)\frac{p\!\!\!\slash_{H}+E_1+E_2}{E_1+E_2}\Gamma_S
(\beta\,p\!\!\!\slash_{H}+k\!\!\!\slash-m_2)\otimes \frac{\bold{1}_c}{\sqrt{N_c}}\,,\label{projection}
\end{eqnarray}
\end{widetext}
where the auxiliary parameter $\omega=\sqrt{E_1+m_b}\sqrt{E_2+m_c}$ and $\bold{1}_c$ is the unit matrix in the fundamental representation of the color SU(3) group. $\Gamma_S=\gamma^5$ for spin-singlet combination with spin $S=0$,
 while $\Gamma_S=\varepsilon\!\!\!\slash_{H}=\varepsilon_\mu(p_H) \gamma^\mu$ for spin-triplet combination with spin $S=1$.

To get the relativistic corrections to the form factors of $B_c$ into $S$-wave charmonium, one may perform the  Taylor expansion of the amplitudes in powers of $k^\mu$
\begin{eqnarray}
{\cal A}(k)&=& {\cal A}(0)+\frac{\partial {\cal A}(0)}{\partial k^\mu}\mid_{k=0}k^\mu\nonumber\\
&&+\frac{1}{2!}\frac{\partial^2 {\cal A}(0)}{\partial k^\mu\partial k^\nu}\mid_{k=0}k^\mu k^\nu+\ldots,
\end{eqnarray}
where the terms linear in $k$ should be dropped since they do not contribute to the matching coefficients.
In this paper,  we will consider the  contributions at the ${\cal O}(|\bold{k}|^2)$ level.  One can use the following replacement to simplify the calculation~\cite{Guo:2011tz}
\begin{eqnarray}
k^\mu k^\nu&\rightarrow& \frac{|\bold{k}|^2}{D-1}(-g^{\mu\nu}+\frac{P_{H}^\mu P_{H}^\nu}{P_{H}^2}).
\end{eqnarray}

The treatment of the final state phase space integrations at the ${\cal O}(|\bold{k}|^2)$ level
is slightly different from the leading order calculation. We will adopt the following rescaling
transformation for the external momenta in order to get their relativistic corrections~\cite{Guo:2011tz}
\begin{eqnarray}
p_H=p_H'\frac{E_1+E_2}{m_1+m_2}.\label{rescale}
\end{eqnarray}

\subsection{Form factors}

The form factors of $B_c$ into $S$-wave charmonium, i.e. $f_{+}$,
$f_{0}$, $V$, $A_{0}$, $A_{1}$, and $A_{2}$  are defined in common~\cite{Wirbel:1985ji}
\begin{eqnarray}
\langle \eta_{c}(p)\vert \bar c \gamma^{\mu}b\vert B_{c}(P)\rangle
&=&f_{+}(q^{2})(P^{\mu}+p^{\mu}-\frac{m_{B_{c}}^{2}-
m_{\eta_{c}}^{2}}{q^{2}}q^{\mu})\nonumber\\ &&+f_{0}(q^{2})
\frac{m_{B_{c}}^{2}-m_{\eta_{c}}^{2}}{q^{2}}q^{\mu}\,,
\end{eqnarray}
\begin{eqnarray}
 && \langle J/\psi(p,\varepsilon^{*})\vert \bar c \gamma^{\mu}b\vert
B_{c}(P)\rangle =\frac{2 i
V(q^{2})}{m_{B_{c}}+m_{J/\psi}}\epsilon^{\mu\nu\rho\sigma}
\varepsilon_{\nu}^{*}p_{\rho}P_{\sigma}\,,
\nonumber\\
&& \langle J/\psi(p,\varepsilon^{*})\vert \bar c
\gamma^{\mu}\gamma_{5}b\vert B_{c}(P)\rangle =2
m_{J/\psi}A_{0}(q^{2})\frac{\varepsilon^{*}\cdot q}{q^{2}}q^{\mu}
\nonumber\\&&~~~~~-A_{2}(q^{2})\frac{\varepsilon^{*}\cdot
q}{m_{B_{c}}+m_{J/\psi}}(
P^{\mu}+p^{\mu}-\frac{m_{B_{c}}^{2}-m_{J/\psi}^{2}}{q^{2}}q^{\mu})
\nonumber\\&&~~~~~+(m_{B_{c}}+m_{J/\psi})A_{1}(q^{2})
(\varepsilon^{*\mu}-\frac{\varepsilon^{*}\cdot q}{q^{2}} q^{\mu})\,,
\end{eqnarray}
where the momentum transfer is defined as $q=P-p$ with the $B_c$ meson momentum $P$ and the charmonium momentum $p$.

\begin{figure}[th]
\begin{center}
\includegraphics[width=0.45\textwidth]{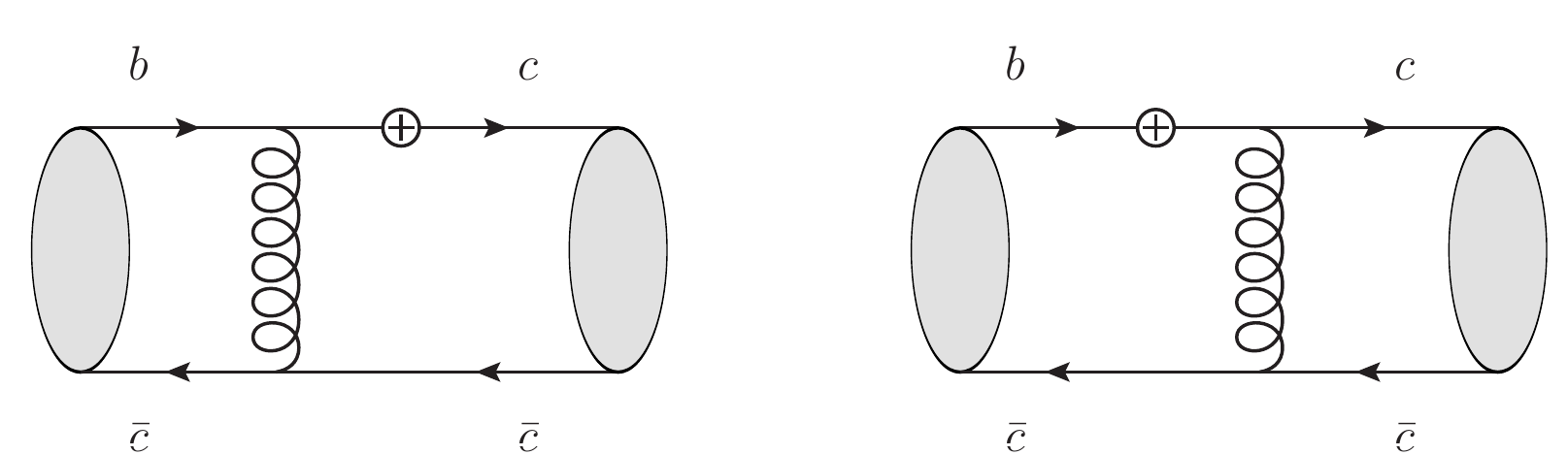}
\end{center}
    \vskip -0.7cm \caption{Feynman diagrams for the form factors of $B_c$ into $S$-wave charmonium, where ``$\oplus$'' denotes certain current operators.}\label{Fig-formfactors}
\end{figure}

The leading order  results at ${\cal O}(\alpha_s )$ and the NLO QCD corrections at ${\cal O}(\alpha_s^2 )$
of the form factors can be found in Refs.~\cite{Bell:2005gw,Qiao:2011yz,Qiao:2012vt}. The leading order  results are obtained
from the Feynman diagrams in Fig.~\ref{Fig-formfactors}. For completeness,
we list the leading order results of the form factors here
\bqa f_+^{LO}(q^{2})&=&8 \sqrt{2} C_A C_F \pi  \sqrt{z+1}
 \alpha _s \psi(0)_{B_c}
\psi(0)_{\eta_c}\nonumber\\&&\times\frac{
\left(-y^2+3 z^2+2 z+3\right) }{\left((1-z)^2-y^2\right)^2 z^{3/2}
m_b^3 N_c} , \eqa
\bqa f_0^{LO}(q^{2})&=&8 \sqrt{2} C_A C_F \pi  \sqrt{z+1}
 \alpha _s
\psi(0)_{B_c}
\psi(0)_{\eta_c} \nonumber\\&&\times\frac{
\left(9 z^3+9 z^2+11 z-y^2 (5 z+3)+3\right) }{\left((1-z)^2-y^2\right)^2 z^{3/2}
(3 z+1) m_b^3 N_c} ,~~
 \eqa
\bqa V^{LO}(q^{2})=\frac{16 \sqrt{2} C_A C_F \pi  (3 z+1) \alpha _s
\psi(0)_{B_c}\psi(0)_{J/\Psi}}{\left((1-z)^2-y^2\right)^2
   \left(\frac{z}{z+1}\right)^{3/2} m_b^3 N_c} ,
 \eqa
\bqa A_0^{LO}(q^2)=\frac{16 \sqrt{2} C_A C_F \pi  (z+1)^{5/2} \alpha
_s\psi(0)_{B_c}\psi(0)_{J/\Psi}}{\left((1-z)^2-y^2
\right)^2 z^{3/2} m_b^3 N_c} ,\nonumber\\
 \eqa
\bqa A_1^{LO}(q^2)&=&16 \sqrt{2} C_A C_F \pi  \sqrt{z+1}  \alpha _s
\psi(0)_{B_c}\psi(0)_{J/\Psi}\nonumber\\&&\times \frac{\left(4
z^3+5 z^2+6 z-y^2(2 z+1)+1\right)}{\left((1-z)^2-y^2\right)^2
z^{3/2} (3 z+1) m_b^3 N_c} ,~~
 \eqa
\bqa A_2^{LO}(q^2)&=&16 \sqrt{2} C_A C_F \pi  \sqrt{z+1}
\alpha_s\psi(0)_{B_c}\psi(0)_{J/\Psi}\nonumber\\&&\times\frac{(3 z+1)
}{\left((1-z)^2-
y^2\right)^2
z^{3/2} m_b^3 N_c} ,
 \eqa
where $z= m_c/m_b$ and $y= \sqrt{q^2/m_b^2}$. $C_A=3$ and $C_F =4/3$, which are the SU(3) color group parameters. The wave functions at the origin of charmonium and $B_c$ meson are defined through the nonperturbative operator matrix elements
\bqa \psi(0)_{\eta_c}&=&\frac{1}{\sqrt{2N_c}}\langle \eta_c| \psi_c^{\dagger}\chi_c|0\rangle,\\
\psi(0)_{B_c}&=&\frac{1}{\sqrt{2N_c}}\langle 0|
\chi_b^{\dagger}\psi_c|B_c\rangle,\\
\psi(0)_{J/\Psi}&=&\frac{1}{\sqrt{2N_c}}\langle J/\Psi| \psi_c^{\dagger}\bfsigma\chi_c|0\rangle.
 \eqa

\subsection{Relativistic corrections}

In the following let us calculate the relativistic corrections to the form factors.
The Feynman diagrams are plotted in Fig.~\ref{Fig-formfactors}. Using the
covariant projection method,
one can get the corresponding amplitudes of Fig.~\ref{Fig-formfactors}. Performing the  Taylor expansion of the amplitudes in powers of $k^\mu$ and extracting the quadratic terms in
the series, one can obtain the relativistic corrections at the ${\cal O}(|\bold{k}|^2)$ level.

The relativistic corrections to the form factors shall be separated into two parts, since there are two bound states, i.e. a charmonium and the $B_c$ meson which are composed of  heavy quark and heavy antiquark.
In the following, let us assign $v$ as the quark relative velocity inside the charmonium and
$v'$ as the equivalent quark relative velocity inside the $B_c$ meson. Then a half of the quark relative momentum is defined as $k=m_c v/2$ inside the charmonium and a half of the quark relative momentum is  defined as $k'=m_{red}v'=m_b m_c v'/(m_b+m_c)$ inside the $B_c$ meson. The masses of the bound states can be written as $m_{\eta_c}=2\sqrt{m_c^2-k^2}$ and
$m_{B_c}=\sqrt{m_c^2-k'^2}+\sqrt{m_b^2-k'^2}$. Using the heavy quark spin symmetry, one can also assume
$m_{J/\psi}=2\sqrt{m_c^2-k^2}$.

If we expand the amplitudes in powers of $k^\mu$, the relativistic corrections from the charm quark-antiquark pair interactions inside the charmonium can be obtained. Analogously, the relativistic corrections to the form factors from the charm and bottom quarks interaction inside the $B_c$ meson can be obtained when we expand the amplitudes in powers of $k'^\mu$.

In order to get the accurate relativistic corrections, one should keep the relative momentum dependence
in the expressions of the masses of charmonium and $B_c$ meson. The rescaling transformation in
Eq.~(\ref{rescale}) for the external charmonium and $B_c$ meson's momenta should be  employed in
order to get the relativistic corrections from the phase space integration.

To estimate the magnitude of the relativistic correction operator matrix elements, we have
\begin{eqnarray}
\langle \eta_c|\psi^\dagger_c \left(-\frac{i}{2}  \overleftrightarrow {\bold D}\right)^2  \chi_c
| 0\rangle&\simeq& |\bold{k}|^2\langle\eta_c|\psi^\dagger_c \chi_c
|0\rangle,  \label{rc1}\\
\langle 0|\chi^\dagger_b \left(-\frac{i}{2}  \overleftrightarrow {\bold D}\right)^2  \psi_c
| B_c\rangle&\simeq& |\bold{k}'|^2\langle 0|\chi^\dagger_b \psi_c
|B_c\rangle,\label{rc2}\\
\langle J/\Psi|\psi^\dagger_c \left(-\frac{i}{2}  \overleftrightarrow {\bold D}\right)^2  \chi_c
| 0\rangle&\simeq& |\bold{k}|^2\langle J/\Psi|\psi^\dagger_c \chi_c
|0\rangle,\label{rc3}
\end{eqnarray}
where $|\bold{k}|^2$ and $|\bold{k}'|^2$ can be described by the heavy quark relative velocities, i.e.
$|\bold{k}|^2=m^2_c |\bold{v}|^2/4$ and $|\bold{k}'|^2=m^2_{red}|\bold{v}'|^2=m^2_b m^2_c |\bold{v}'|^2/(m_b+m_c)^2$.

First let us keep the charm quark relative momentum $k^\mu$ nonzero in the charmonium and set the quark relative momentum $k'^\mu=0$  in the $B_c$ meson. Expanding the amplitudes in powers of $k^\mu$, we can
obtain the related relativistic corrections.
The relativistic corrections can be expressed as the product of the relativistic operator's matrix elements
and the related short-distance coefficients. According to Eqs.~(\ref{rc1}-\ref{rc3}), the relativistic operator's matrix elements can be estimated by the wave functions at the origin of heavy quarkonium.
The results of relativistic corrections from the charmonium can be written as  the following simple forms

\begin{widetext}
\bqa f_+^{RC}(q^{2})&=&|\bold{k}|^2 f_+^{LO}(q^2)\frac{-3 y^4+8 y^2 z (2 z-3)-53 z^4+4 z^3-6 z^2+52 z+3}{4 m_b^2 z^2 \left((z-1)^2-y^2\right) \left(-y^2+3 z^2+2 z+3\right)} , \label{rcfp} \eqa
\bqa f_0^{RC}(q^{2})&=&|\bold{k}|^2 f_0^{LO}(q^2)\left(-\frac{\left(-3 z^2+2 z+1\right)^2 \left(53 z^3+49 z^2+55 z+3\right)}{4 m_b^2 (z-1) z^2 (3 z+1) \left((z-1)^2-y^2\right) \left(-y^2 (5 z+3)+9 z^3+9 z^2+11
   z+3\right)}\right.\nonumber\\&&\left.+\frac{y^2 \left(y^4 (3 z+1)-y^2 \left(83 z^3+55 z^2+17 z+5\right)+365 z^5-173 z^4-14 z^3+246 z^2+81 z+7\right)}{4 m_b^2 (z-1) z^2 (3 z+1)
   \left((z-1)^2-y^2\right) \left(-y^2 (5 z+3)+9 z^3+9 z^2+11 z+3\right)}\right) ,~~
 \eqa
\bqa V^{RC}(q^{2})&=&|\bold{k}|^2 V^{LO}(q^2)\frac{-y^2 \left(24 z^2+27 z+5\right)-12 z^4+87 z^3+171 z^2+69 z+5}{6 m_b^2 z^2 (z+1) (3 z+1) \left((z-1)^2-y^2\right)} ,
 \eqa
\bqa A_0^{RC}(q^2)&=&|\bold{k}|^2 A_0^{LO}(q^2)\frac{-3 y^4-2 y^2 \left(14 z^3+5 z^2-3\right)-4 z^5+85 z^4+348 z^3+214 z^2-3}{24 m_b^2 z^3 (z+1)^2 \left((z-1)^2-y^2\right)}\; ,
 \eqa
\bqa A_1^{RC}(q^2)&=&|\bold{k}|^2 A_1^{LO}(q^2)\left(\frac{-45 z^6+721 z^5+1554 z^4+1954 z^3+807 z^2+125 z+4}{12 m_b^2 z^2 (3 z+1) \left((z-1)^2-y^2\right) \left(-y^2 (2 z+1)+4 z^3+5 z^2+6 z+1\right)}\right.\nonumber\\&&\left.-\frac{y^2 \left(y^2 \left(3 z^2-7 z-4\right)+48 z^4+580 z^3+512 z^2+132 z+8\right)}{12 m_b^2 z^2 (3 z+1) \left((z-1)^2-y^2\right) \left(-y^2 (2 z+1)+4 z^3+5
   z^2+6 z+1\right)}\right) ,~~
 \eqa

\bqa A_2^{RC}(q^2)&=&|\bold{k}|^2 A_2^{LO}(q^2)\frac{-y^2 \left(39 z^2+55 z+10\right)+15 z^4-79 z^3+187 z^2+123 z+10}{12 m_b^2 z^2 (3 z+1) \left((z-1)^2-y^2\right)} .\label{rca2}
 \eqa
 \end{widetext}
According to the leading order results, we have learned that the form factors have singularity at the minimum recoil point $y\to 1-z$, which exist also in the relativistic corrections from Eqs.~(\ref{rcfp}-\ref{rca2}).

Because $z=
m_c/m_b\approx 0.3$  is small, the form factors can be expanded in powers of $z$. In the heavy quark limit  $m_b \to \infty$, one can get more information among form factors.
In the $m_b \to \infty$ limit, the form factors become
 \bqa
V(q^{2})_{m_b \rightarrow\infty}&=&\frac{16 \sqrt{2} C_A C_F \pi
\alpha _s
\psi(0)_{B_c}\psi(0)_{J/\Psi}}{\left(1-y^2\right)^2
   z^{3/2} m_b^3 N_c}\nonumber\\&&\left(1+\frac{5|\bold{k}|^2}{6m_b^2z^2}\right)\; ,
 \eqa
\bqa A_2(q^2)_{m_b \rightarrow\infty}
=V(q^2)_{m_b \rightarrow\infty}\; ,
 \eqa
 \bqa A_0(q^2)^{LO}_{m_b \rightarrow\infty}
=V(q^2)^{LO}_{m_b \rightarrow\infty}\; .
 \eqa

At the maximum recoil  point with $q^2=0$ , some form factors turn to be identical
 \bqa f_0(0)=f_+(0)\;\label{ff1} ,
 \eqa
\bqa V(0)_{m_b \rightarrow\infty}=A_2(0)_{m_b \rightarrow\infty}\; ,
 \eqa
which are consistent with the predictions of the heavy quark effect theory~\cite{Stech:1995ec}
and the large energy effective theory~\cite{Charles:1998dr}.

Next let us keep the quark relative momentum $k'^\mu$ nonzero in the $B_c$ meson and set  the charm quark relative momentum $k^\mu=0$  in the charmonium. We will expand the amplitudes in powers of $k'^\mu$, and the relativistic corrections to the form factors at the ${\cal O}(|\bold{k}'|^2)$ level are
\begin{widetext}
\bqa f_+^{RC'}(q^{2})&=&|\bold{k}'|^2 f_+^{LO}(q^2)\left(\frac{-9 z^6+264 z^5+285 z^4+408 z^3+241 z^2+112 z-21}{24 z^2 m_b^2 \left((z-1)^2-y^2\right) \left(-y^2+3 z^2+2 z+3\right)}\right.\nonumber\\&&\left.-\frac{y^2 \left(y^2 \left(3 z^2-4 z+11\right)+4 \left(-3 z^4+32 z^3+23 z^2+16 z-8\right)\right)}{24 z^2 m_b^2 \left((z-1)^2-y^2\right) \left(-y^2+3 z^2+2
   z+3\right)}\right) , \eqa
\bqa f_0^{RC'}(q^{2})&=&|\bold{k}'|^2 f_0^{LO}(q^2)\left(-\frac{(3 z+1) \left(9 z^7-273 z^6-21 z^5-123 z^4+167 z^3+129 z^2+133 z-21\right)}{24 (z-1) z^2 m_b^2 \left((z-1)^2-y^2\right) \left(-y^2 (5 z+3)+9 z^3+9
   z^2+11 z+3\right)}\right.\nonumber\\&&\left.+\frac{y^2 \left(63 z^7-1110 z^6-762 z^5-189 z^4+307 z^3+216 z^2-40 z-21\right)}{12 (z-1) z^2 (3 z+1) m_b^2 \left((z-1)^2-y^2\right) \left(-y^2 (5 z+3)+9 z^3+9
   z^2+11 z+3\right)}\right.\nonumber\\&&\left.+\frac{y^4 \left(\left(32-6 y^2\right) z^2+\left(31-2 y^2\right) z-15 z^5+161 z^4+104 z^3+7\right)}{8 (z-1) z^2 (3 z+1) m_b^2 \left((z-1)^2-y^2\right)
   \left(-y^2 (5 z+3)+9 z^3+9 z^2+11 z+3\right)}\right) ,~~
 \eqa
\bqa V^{RC'}(q^{2})&=&|\bold{k}'|^2 V^{LO}(q^2)\frac{y^2 \left(9 z^3-105 z^2-23 z+7\right)-9 z^5+315 z^4+108 z^3+180 z^2+53 z-7}{24 z^2 (3 z+1) m_b^2 \left((z-1)^2-y^2\right)} ,
 \eqa
\bqa A_0^{RC'}(q^2)&=&|\bold{k}'|^2 A_0^{LO}(q^2)\left(\frac{-12 z^7+277 z^6+916 z^5+605 z^4+596 z^3+207 z^2-28 z-1}{96 z^3 (z+1)^2 m_b^2 \left((z-1)^2-y^2\right)}\right.\nonumber\\&&\left.+\frac{2 y^2 \left(6 z^5+5 z^4-140 z^3-22 z^2+22 z+1\right)-y^4 \left(39 z^2+16 z+1\right)}{96 z^3 (z+1)^2 m_b^2 \left((z-1)^2-y^2\right)}\right)\; ,
 \eqa
\bqa A_1^{RC'}(q^2)&=&|\bold{k}'|^2 A_1^{LO}(q^2)\left(\frac{-36 z^8+807 z^7+2047 z^6+2807 z^5+2503 z^4+1693 z^3+421 z^2+5 z-7}{24 z^2 (3 z+1) m_b^2 \left((z-1)^2-y^2\right) \left(-y^2 (2 z+1)+4 z^3+5 z^2+6 z+1\right)}\right.\nonumber\\&&\left.-\frac{ y^2 \left(18 z^4-33 z^3-9 z^2+25 z+7\right)+2 \left(-27 z^6+258 z^5+539 z^4+504 z^3+151 z^2-10 z-7\right)}{24 z^2 y^{-2}(3 z+1) m_b^2
   \left((z-1)^2-y^2\right) \left(-y^2 (2 z+1)+4 z^3+5 z^2+6 z+1\right)}\right) ,~~~
 \eqa

\bqa A_2^{RC'}(q^2)&=&|\bold{k}'|^2 A_2^{LO}(q^2)\frac{-y^2 \left(12 z^3+67 z^2+16 z-3\right)-6 z^5+175 z^4+102 z^3+92 z^2+24 z-3}{12 z^2 (3 z+1) m_b^2 \left((z-1)^2-y^2\right)} .
 \eqa

\begin{table}
\begin{center}
\caption{$B_c$ into $S$-wave charmonium form factors at
$q^2=0$ evaluated in the literatures.
}
\begin{tabular}{c|c|cccccccccc}
 \hline\hline
   \multicolumn{2}{c|}{Approaches}& $f_+^{B_c \eta_c}(0)=f_0^{B_c \eta_c}(0)$~~ & $A_0^{B_c J/\psi}(0)$~~ & $A_1^{B_c J/\psi}(0)$~~
    & $A_2^{B_c J/\psi}(0)$  ~~ & $V^{B_c J/\psi}$(0) \\
 \hline
 \multirow{4}{*}{PQCD}&DW\cite{Du:1988ws}\footnote{We quote the results with $\omega=0.6$ GeV.}
    & $0.420$ & $0.408$ & $0.416$ & $0.431$  & $0.296$ \\
\cline{2-7}
 & SDY\cite{Sun:2008ew}
     & $0.87$ &  $0.27$ &$0.75$  & $1.69$ & $0.85$ \\
\cline{2-7}
    &WFX\cite{Wen-Fei:2013uea}
     & $0.48$  & $0.59$ &$0.46$  & $0.64$  & $0.42$ \\
 \cline{2-7}
   &ZLWX\cite{Rui:2016opu}
     & $1.06$  & $0.78$ &$0.96$  & $1.36$  & $1.59$ \\\hline
     \multirow{3}{*}{QCD SR}&CNP\cite{Colangelo:1992cx}
     & $0.20$  & $0.26$ &$0.27$  & $0.28$  & $0.19$\\
  \cline{2-7}
   & KT\cite{Kiselev:1993ea}
     & $0.23$  & $0.21$ &$0.21$  & $0.23$  & $0.17$ \\
  \cline{2-7}
    &KLO\cite{Kiselev:1999sc}
     & $0.66$  & $0.60$ &$0.63$  & $0.69$  & $0.52$ \\
  \hline
   LCSR& HZ\cite{Huang:2007kb}
     & $0.87$ &  $0.27$ &$0.75$  & $1.69$ & $1.69$ \\
 \hline
   \multirow{3}{*}{RQM} & NW\cite{Nobes:2000pm}
     & $0.5359$ & $0.532$ &$0.524$  &$0.509$ & $0.368$ \\
   \cline{2-7}
   & EFG\cite{Ebert:2003cn}
     & $0.47$ & $0.40$  &$0.50$  & $0.73$ & $0.25$ \\
   \cline{2-7}
   & IKS2\cite{Ivanov:2005fd}
     & $0.61$ & $0.57$  &$0.56$  & $0.54$ & $0.42$ \\
    \hline
    NCQM&HNV\cite{Hernandez:2006gt}
     & $0.49$ & $0.45$  &$0.49$  & $0.56$ & $0.31$\\
     \hline
  \multirow{2}{*}{LFQM}& WSL\cite{Wang:2008xt}
     & $0.61$ &  $0.53$ &$0.50$  & $0.44$ & $0.37$ \\\cline{2-7}
   & KLL\cite{Ke:2013yka}
     & -- &  $0.502$ &$0.467$  & $0.398$ & $0.369$ \\\hline
      PMC&SWMW\cite{Shen:2014msa}
     & $1.65$ & $0.87$  &$1.07$  & $1.15$ & $1.47$\\\hline
      \multirow{3}{*}{NRQCD}& LO~\cite{Bell:2005gw,Qiao:2011yz,Qiao:2012vt}
     & $0.96$ &  $0.84$ &$0.87$  & $0.94$ & $1.21$ \\\cline{2-7}
   & NLO~\cite{Bell:2005gw,Qiao:2011yz,Qiao:2012vt}
     & $1.43$ &  $1.09$ &$1.19$  & $1.27$ & $1.63$ \\
     \cline{2-7}
   & NLO+RC(This work)
     & $1.67$ &  $1.43$ &$1.57$  & $1.73$ & $2.24$ \\
    \hline\hline
\end{tabular}\label{table1}
\end{center}
\end{table}

\begin{table}
\begin{center}
\caption{Comparisons of the results of the decay ratios of $B_c \to J/\psi+D_s^{(*)}$ with data and other theoretical predictions.
\label{br}}
\begin{tabular}{cccccc}
\hline\hline
$R_{D_s^+/\pi^+}$ & $R_{D_s^{*+}/\pi^+}$ &   $R_{D_s^{*+}/D_s^+}$&   $\Gamma_{\pm\pm}/\Gamma$ &~~~ Refs.\\
\hline
$3.8\pm1.2$ & $10.4\pm3.5$ &$2.8^{+1.2}_{-0.9}$ & $0.38\pm0.24$ & ATLAS\cite{Aad:2015eza}\\
$2.90\pm0.62$ ~~& -- &~~$2.37\pm0.57$ & ~~$0.52\pm0.20$ & LHCb\cite{Aaij:2013gia}.\\
2.6 ~~& 4.5 &~~1.7 & ~~--& Potential model\cite{Colangelo:1999zn}\\
1.3 ~~& 5.2 &~~3.9 & ~~--& QCD SR\cite{Kiselev:2002vz}\\
2.0 ~~& 5.7 &~~2.9 & ~~--& RCQM\cite{Ivanov:2006ni}\\
2.2 ~~& -- &~~-- & ~~--& BSW\cite{Dhir:2008hh}\\
$2.06\pm0.86$ ~~& -- &~~$3.01\pm1.23$ & ~~--& LFQM\cite{Ke:2013yka}\\
$3.45^{+0.49}_{-0.17}$ ~~& -- &~~$2.54^{+0.07}_{-0.21}$ & ~~$0.48\pm0.04$&PQCD\cite{Rui:2014tpa}\\
-- ~~& -- &~-- & ~~0.410& RIQM\cite{Kar:2013fna}\\
$3.07^{+0.21+0.14}_{-0.38-0.13}$ & $11.8^{+1.0+2.3}_{-1.4-0}$ &$3.85^{+0.04+0.54}_{-0.02-0}$ & $0.601^{+0.001+0.033}_{-0.001-0.040}$ & ~~NRQCD NLO+RC\\
\hline\hline
\end{tabular}
\end{center}
\end{table}

 \end{widetext}

In the heavy quark limit $m_b \to \infty$, however, the relations among the relativistic corrections
to the form factors from the $B_c$ meson have not been found.

In order to estimate the relativistic corrections, one should first  evaluate  the heavy quark relative
velocities. Using the heavy quark kinetic and potential energy approximation~\cite{Bodwin:1994jh}, one has
\begin{eqnarray} |\mathbf{  v}| &\simeq&\alpha_s(2m_{red}|\mathbf{  v}| \,)\,.
\end{eqnarray}
We adopt the values which have been evaluated in Ref.~\cite{Wang:2015bka}
\begin{eqnarray}
| \mathbf{v}|^2
_{J/\psi}&\approx&0.267\,,~~~~~~
| \mathbf{v}'|^2
_{B_c}\approx0.186\,.
\end{eqnarray}

In the literatures, there already exist a lot of studies on the form factors of $B_c$ into $S$-wave
charmonium at the maximum recoil point, so we give the results with different approaches in Tab.~\ref{table1}.
For the NRQCD predictions in Tab.~\ref{table1}, the heavy quark masses $m_c=1.4$ and $m_b=4.9$ are adopted. From the table,
the NRQCD predictions of the form factors are larger than that from the approaches of QCD SR, RQM, NCQM and LFQM.
The predictions of some of form factors are consistent with each other among  PQCD, LCSR, PMC and the LO NRQCD
results. The relativistic corrections from both the $S$-wave charmonium and $B_c$ can not be ignored, which will bring about
15 to 25 percent enhancements to the NLO predictions of the form factors.

\section{Decay ratios of $B_c \to J/\psi+D_s^{(*)}$\label{III}}

The form factors of  $B_c$ into $S$-wave charmonium  can be employed into the calculation of decay widths of
a lot of decay channels, some of which have been studied by the LHCb and ATLAS Collaborations.
The channels of $B_c$ exclusive decays to a charmonium and  a light meson can determine the form factors
at the maximum recoil point, while the exclusive decays to a charmonium and  a heavy meson can give the
information of the form factors far from the  maximum recoil point.

The ATLAS Collaboration have used a dataset of integrated luminosities of 4.9$fb^{-1}$ and $20.6 fb^{-1}$
at $\sqrt{s}=$ 7TeV and 8TeV  and measured the branching ratios of  $B^+_c \to J/\psi+D_s^{+(*)}$ recently
~\cite{Aad:2015eza}. Furthermore, the  ATLAS Collaboration have analyzed three helicity dependent amplitudes
in the channel of  $B^+_c \to J/\psi+D_s^{+(*)}$, i.e. $A_{++}$, $A_{--}$, and $A_{00}$, where the subscripts
correspond to the helicities of $J/\psi$ and $D_s^{*}$ mesons. Therein $A_{++}$ and $A_{--}$ denote
the amplitudes where $J/\psi$ and $D_s^{*}$ are transversely polarized.
We will employ the form factors to analyze these channels, and compare our results with data and other theoretical predictions.

\begin{figure}[th]
\begin{center}
\includegraphics[width=0.42\textwidth]{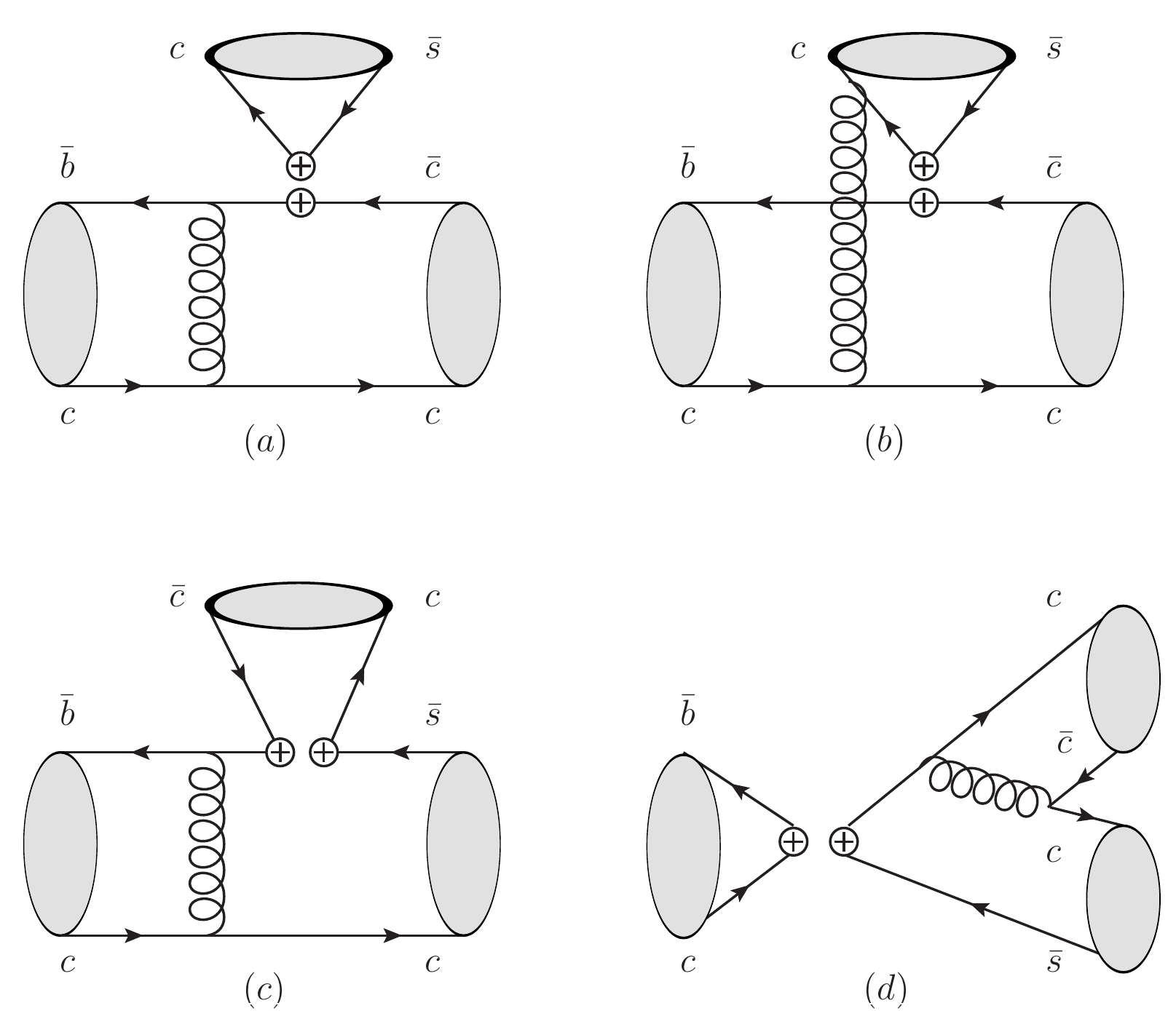}
\end{center}
    \vskip -0.7cm \caption{Typical Feynman diagrams for $B^+_c \to J/\psi+D_s^{+(*)}$, where two ``$\oplus$'' denote four-fermion weak interaction operators. There are four types of topologies: (a) factorizable diagrams; (b)non-factorizable diagrams; (c)color-suppressed diagrams; (d) annihilation diagrams.}\label{Fig-Ds}
\end{figure}

The typical Feynman diagrams of  $B^+_c \to J/\psi+D_s^{+(*)}$ are plotted in Fig.~\ref{Fig-Ds}. There are four
types of topologies:
(a) factorizable diagrams which are determined by the form factors;
(b) non-factorizable diagrams which can not be factorized into the product of form factors and
the decay constant of the $D_s^{(*)}$ meson;
(c) color-suppressed diagrams where the spectator quark generates the $D_s^{(*)}$ meson with a strange quark;
(d) annihilation diagrams where both the bottom and anti-charm quarks in the $B_c$ meson are annihilated.

The form factors dependence on $q^2$  are obtained using the NRQCD approach, however, a divergence exists at the minimum
recoil point where $q^2=(m_{B_c}-m_{J/\psi})^2\approx(m_b-m_c)^2$ according to the LO results~\cite{Bell:2005gw}
and the NLO QCD corrections~\cite{Bell:2005gw,Qiao:2011yz,Qiao:2012vt} and relativistic corrections.
This singularity problem leads to the predictions of the form factors invalid near the minimum recoil point.
Besides, the NRQCD prediction of the branching ratio of $B^+_c \to J/\psi+\pi^++\pi^-+\pi^+$
is slightly larger than the cental value of the measurement by the LHCb Collaboration~\cite{LHCb:2012ag,Qiao:2012hp},
where the channel  $B_{c}^{+}\to J/\psi a_1^+(1260)$  dominates the contribution. This indicates the applications of
the NRQCD predictions of the form factors far from the maximum recoil point should be careful.
For the channels  $B^+_c \to J/\psi+D_s^{+(*)}$ where $q^2=m_{D_s^{(*)}}^2\approx 4 {GeV}^2$,
the direct NRQCD predictions of form factors maybe invalid, since $q^2 \approx 4 {GeV}^2$ is far
from the maximum recoil point.

To extrapolate the form factors to the minimum recoil region,
the pole mass dependence model are generally adopted in many literatures~\cite{Kiselev:1999sc,Wang:2008xt}, where each form factor $F(q^2)$ is
parametrized as
\begin{equation}\label{pole mass}
    F(q^2)=\frac{F(0)}{1-\frac{q^2}{m^2_{\mathrm{pole}}}-
    \beta\frac{ q^4}{m^4_{\mathrm{pole}}}}\; ,
\end{equation}
with  the effective pole mass $m_{\mathrm{pole}}$ and a free parameter $\beta$ which is set to be zero in our
calculation. And $F(q^2)$ can be any one of the form factors of $B_c$ into $S$-wave charmonium, i.e. $f_{+}(q^2)$,
$f_{0}(q^2)$, $V(q^2)$, $A_{0}(q^2)$, $A_{1}(q^2)$, and $A_{2}(q^2)$.

In the calculation, the heavy quark mass are adopted as $m_c=1.4\pm0.1\mathrm{GeV}$, $m_b=4.9\pm0.1\mathrm{GeV}$~\cite{Qiao:2012hp,Zhu:2015jha}.
The masses of $D_s^{(*)}$ are adopted as $m_{D_s}=1.968$GeV and $m_{D_s^*}=2.112$GeV~\cite{Olive:2016xmw}.
The decay constants are adopted as $f_\pi=130.4\mathrm{MeV}$~\cite{Qiao:2012hp}, $f_{D_s}=257.5\mathrm{MeV}$~\cite{Rui:2014tpa}. The decay constant of $D_s^*$
can be obtained by the heavy quark symmetry $f_{D_s^*}=f_{D_s}(m_{D_s}/m_{D_s^*})^{1/2}$.
The effective pole mass $m_{\mathrm{pole}}$ in Eq.~(\ref{pole mass}) is set to be near the bottom quark mass, i.e. 5GeV.
 The Schr\"{o}dinger wave function
at the origin for  $J/\psi$ is determined through its leptonic decay
width $\Gamma_{ee}^\psi=5.55\mathrm{keV}$.
Numerically we can obtain  $|\psi
_{\Psi}^{LO}(0)|^2=0.0447(\mathrm{GeV})^3$ and $|\psi
_{\Psi}^{NLO}(0)|^2=0.0801(\mathrm{GeV})^3$. For that of $B_c$, we
shall determine its value to be:
$|\psi_{B_c}(0)|^2=0.1307(\mathrm{GeV})^3$, which is derived under
the Buchm\"{u}ller-Tye potential\cite{Eichten:1994}.

Based on the NRQCD framework, we can calculate the amplitudes in Fig.~\ref{Fig-Ds} and numerical results indicate that  the
factorizable diagrams dominate the contribution of the decay widths of  $B^+_c \to J/\psi+D_s^{+(*)}$, but colour-suppressed and annihilation topologies diagrams contribute less than 10 percent.  The
factorizable diagrams can be factorized into the form factor part and the $D_s^{(*)}$ decay constant part. Thus we can employ the results of  NLO QCD and relativistic corrections to the form factors, and obtain more
precise predictions.

 In order to compare with data, the auxiliary parameters are written as
\begin{eqnarray}
R_{D_s^+/\pi^+}&=&\frac{\Gamma(B^+_c \to J/\psi+D_s^{+})}{\Gamma(B^+_c \to J/\psi+\pi^{+})} ,\\
R_{D_s^{*+}/\pi^+}&=&\frac{\Gamma(B^+_c \to J/\psi+D_s^{*+})}{\Gamma(B^+_c \to J/\psi+\pi^{+})},
\end{eqnarray}
\begin{eqnarray}
R_{D_s^{*+}/D_s^+}&=&\frac{\Gamma(B^+_c \to J/\psi+D_s^{*+})}{\Gamma(B^+_c \to J/\psi+D_s^+)},\\
\Gamma_{\pm\pm}/\Gamma&=&\frac{\Gamma_{\pm\pm}(B^+_c \to J/\psi+D_s^{*+})}{\Gamma(B^+_c \to J/\psi+D_s^{*+})}.
\end{eqnarray}

The decay width  can be written as
\begin{eqnarray}
\Gamma(B_c\to J/\psi D_s^{(*)})&=&\frac{|\mathbf{p}|}{8\pi
m_{B_c}^2}|{\cal A}(B_{c}\to J/\psi D_s^{(*)})|^2,~~~~
\end{eqnarray}
with the final meson momentum  $|\mathbf{p}|=(m_{B_c}^4-2m_{B_c}^2(m_{D_s^{(*)}}^2+m_{\psi}^2)+(m_{D_s^{(*)}}^2-m_{\psi}^2)^2)^{1/2}/(2m_{B_c})$ in the $B_c$ meson rest frame.

Ignoring the small contributions from non-factorizable, color-suppressed and annihilation diagrams,
we reach the naive factorization. In naive factorization, the decay amplitudes can be factorized as
\begin{eqnarray}
&&{\cal A}(B^+_{c}\to J/\psi D_s^+)\nonumber\\&\approx&\frac{G_F}{\sqrt{2}}V_{cb}^*V_{cs}C_{0}(\mu)
\langle J/\psi D_s^{+}\vert {\cal O}_{0}\vert B_{c}^{+}\rangle\nonumber\\
&\approx&\frac{G_F}{\sqrt{2}}V_{cb}^*V_{cs}C_{0}(\mu)\langle J/\psi\vert \bar b
\gamma^{\mu}(1-\gamma_{5})c\vert B_{c}^{+}\rangle
\nonumber\\
&&\times \langle
D_s^{+}\vert \bar c\gamma_{\mu}(1-\gamma_{5})s\vert 0\rangle
\nonumber\\
&=& i\frac{2G_F}{\sqrt{2}}V_{cb}^*V_{cs}C_{0}(\mu)
f_{D_s} A_{0}(m_{D_s}^2)m_{J/\psi}\varepsilon_{J/\psi}^*\cdot q \,,~~~~\label{Ds}
\end{eqnarray}
where $G_F$ is the Fermi constant.
$V_{ud}$ and $V_{cb}$ are the
Cabibbo-Kobayashi-Maskawa (CKM) matrix-elements. ${\cal O}_{0}$ is
the effective four-quark operator, and $C_{0}(\mu)$ is the
perturbatively calculable Wilson coefficient. $\varepsilon_{J/\psi}$ is
the polarization vector of $J/\psi$.
Because $\varepsilon_{J/\psi}^*\cdot q=\varepsilon_{J/\psi}^*\cdot P_{B_c}$ is nonzero
only when $\varepsilon_{J/\psi}^\mu$ is longitudinal,   $J/\psi$ is longitudinally polarized
in the decay channels of $B_c$ exclusive decays to $J/\psi$ and  a pseudoscalar meson. In order to
reliably predict the decay rate of $B_c$ exclusive decays to $J/\psi$ and  a heavy meson, we will adopt the
pole mass dependence model and get $A_{0}(m_{D_s}^2)=A^{\mathrm{NLO} +\mathrm{RC}}_{0}(0)
/(1-\frac{m_{D_s}^2}{m^2_{\mathrm{pole}}})$.

Analogously, we can obtain the expression of the amplitude
${\cal A}(B^+_{c}\to J/\psi \pi^+)$ by replacing the
meson decay constant $f_{D_s}\to f_\pi$, the form factors
$A_{0}(m_{D_s}^2)\to A_{0}(m_{\pi}^2)$  and the CKM matrix-element $V_{cs}\to V_{ud}$ in Eq.~(\ref{Ds}).
Because the pion's mass is much less than
the heavy quark mass, the pion's mass can be ignored in the decay of $B^+_{c}\to J/\psi \pi^+$.

The amplitude of $B^+_{c}\to J/\psi D_s^{*+}$ can be  estimated as
\begin{eqnarray}
&&{\cal A}(B^+_{c}\to J/\psi D_s^{*+})\nonumber\\&\approx&\frac{G_F}{\sqrt{2}}V_{cb}^*V_{cs}C_{0}(\mu)\langle J/\psi D_s^{*+}\vert {\cal O}_{0}\vert B_{c}^{+}\rangle\nonumber\\
&=& i\frac{G_F}{\sqrt{2}}V_{cb}^*V_{cs}C_{0}(\mu)
f_{D^*_s} \varepsilon_{J/\psi}^{*\alpha} \varepsilon_{D^*_s}^{*\beta} \left(
S_1 g_{\alpha\beta}\right.
\nonumber\\&&\left.-S_2\frac{P_{B_c \alpha}P_{B_c \beta}}{m_{B_c}^2}+i S_3\epsilon_{\alpha\beta\gamma\sigma}\frac{P_{J/\psi}^\gamma P_{D_s}^\sigma}{P_{J/\psi}\cdot P_{D_s}}\right) \,,~~~~\label{Dss}
\end{eqnarray}
where
\begin{eqnarray}
S_1&=&-m_{D^*_s}A_1(m_{D^*_s}^2)(m_{B_c}+m_{J/\psi}),\nonumber\\
S_2&=&-\frac{2 m_{D^*_s}A_2(m_{D^*_s}^2)m_{B_c}^2}{m_{B_c}+m_{J/\psi}},\nonumber\\
S_3&=&\frac{2 m_{D^*_s}V(m_{D^*_s}^2)P_{J/\psi}\cdot P_{D_s}}{m_{B_c}+m_{J/\psi}}.
\end{eqnarray}
In the $B_c$ meson rest frame, it is convenient to choose the momentum $\mathbf{P}_{D^*_s}$ to be directed in positive $z$-direction. The transverse polarization vectors  can be defined as $\varepsilon_{D^*_s,\pm}^\mu=\varepsilon_{J/\psi,\mp}^\mu=(0,\pm1,i,0)/\sqrt{2}$. The longitudinal  polarization vectors  can be defined as $\varepsilon_{D^*_s,0}^\mu=(\mathbf{P}_{D^*_s},0,0,P^0_{D^*_s})/m_{D^*_s}$ and $\varepsilon_{J/\psi,0}^\mu=(-\mathbf{P}_{D^*_s},0,0,P^0_{J/\psi})/m_{J/\psi}$. We then can get
the results for the polarized decay widths $\Gamma_{\pm\pm}(B^+_c \to J/\psi+D_s^{*+})$.

Considering the NLO QCD corrections and the relativistic corrections, our results are given in the end
line of Tab.~\ref{br}. For convenience,  we also list the data and other theoretical predictions in Tab.~\ref{br}.
For our results, the first column of the uncertainties is from the choice of the scale $\mu=4.9\pm1$GeV,
while the second error is from the uncertainty of the heavy quark mass with
$m_c=1.4\pm0.1\mathrm{GeV}$ and $m_b=4.9\pm0.1\mathrm{GeV}$.

From  Tab.~\ref{br}, our results of $R_{D_s^+/\pi^+}$, $R_{D_s^{*+}/\pi^+}$,
and $\Gamma_{\pm\pm}/\Gamma$ are consistent with the LHCb and ATLAS data when considering the experiment uncertainties.

\section{Conclusion}

In this paper, we calculated the relativistic corrections to the form factors of $B_c$ into $S$-wave charmonium
at the ${\cal O}(|\bold{k}|^2)$  and  ${\cal O}(|\bold{k}'|^2)$ level, where
$k$ and $k'$ are a half of quark relative momentum inside the charmonium and  $B_c$ meson, respectively.
The corresponding analytic expression are given. In the heavy bottom quark limit,
the properties of form factors are studied.  We found that the relativistic corrections  bring about
15 to 25 percent enhancements to the form factors.

Based on the NRQCD approach, we studied the decay channels of  $B_c \to J/\psi D_s^{(*)}$.  Employing the
form factors of $B_c$ meson into $S$-wave charmonium up to the next-to-leading order in both  $\alpha_s$
and the quark relative velocity squared $v^2$ and $v'^2$,  the decay rates of  $B^+_c \to J/\psi D_s^{(*)+}$
are studied.  Numerical results indicate that the factorizable diagrams dominate the decay rates
of the considered $B_c \to J/\psi D_s^{(*)}$ decay modes.
The ratios of $R_{D_s^+/\pi^+}$, $R_{D_s^{*+}/\pi^+}$, $R_{D_s^{*+}/D_s^+}$ and $\Gamma_{\pm\pm}/\Gamma$
provide a precise platform to test the form factors. Our results of $R_{D_s^+/\pi^+}$, $R_{D_s^{*+}/\pi^+}$,
and $\Gamma_{\pm\pm}/\Gamma$ are consistent with the LHCb and ATLAS data, however, the result
of $R_{D_s^{*+}/D_s^+}$ only support the ATLAS data. Thus more studies  are needed to investigate
the inner properties of $D_s^{*}$.  This work is also helpful to understand the nonperturbative
properties of heavy quarkonium.

\section*{Acknowledgments}
This work was supported in part by the National Natural Science Foundation
of China under Grant No. 11647163 and 11235005, and by the Research Start-up Funding (R.L. Zhu) of
Nanjing Normal University.

\end{document}